\documentstyle[12pt]{article}
\newcommand{\bce}{\begin{center}}
\newcommand{\ece}{\end{center}}
\newcommand{\beq}{\begin{equation}}
\newcommand{\eeq}{\end{equation}}
\newcommand{\bea}{\vspace{0.25cm}\begin{eqnarray}}
\newcommand{\eea}{\end{eqnarray}}

\newcommand{\ba}{\begin{array}}
\newcommand{\ea}{\end{array}}

\newcommand{\doublespace}{
    \renewcommand{\baselinestretch}{1.6}\large\normalsize}

\def\lsim{\mathrel{\rlap{\lower4pt\hbox{\hskip1pt$\sim$}}
    \raise1pt\hbox{$<$}}}         
\def\gsim{\mathrel{\rlap{\lower4pt\hbox{\hskip1pt$\sim$}}
    \raise1pt\hbox{$>$}}}         

\def\Pom{{\bf I\!P}}

\def\lsim{\mathrel{\rlap{\lower4pt\hbox{\hskip1pt$\sim$}}
    \raise1pt\hbox{$<$}}}         
\def\gsim{\mathrel{\rlap{\lower4pt\hbox{\hskip1pt$\sim$}}
    \raise1pt\hbox{$>$}}}         

\def\Pom{{\bf I\!P}}

  \advance\oddsidemargin  by -1in
  \advance\evensidemargin by -1in
  \advance\topmargin      by -0.5in
\def\lsim{\mathrel{\rlap{\lower4pt\hbox{\hskip1pt$\sim$}}
    \raise1pt\hbox{$<$}}}         
\def\gsim{\mathrel{\rlap{\lower4pt\hbox{\hskip1pt$\sim$}}
    \raise1pt\hbox{$>$}}}         

\def\Pom{{\bf I\!P}}
\def\beq{\begin{equation}}
\def\endeq{\end{equation}}
\def\arr{\begin{eqnarray}}
\def\endarr{\end{eqnarray}}
\makeindex

\begin{document}

\pagestyle{empty}
\phantom.\hspace{8.8cm}{\Large\vbox{
\hbox{\bf KFA-IKP(TH)-1994-43}
\hbox{\bf ADP-94-28/T167} } }
\bigskip\\
\phantom. \hspace{10.2cm}
\vspace{1.5cm}\\
\begin{center}
{\bf \huge
Quantum coherence in heavy-flavor production on nuclei
\vspace{1.0cm}}

{\Large N. N. Nikolaev$^{a,b)}$, G. Piller$^{c)}$ and B. G. Zakharov$^{b)}$
\medskip\\ }
{\large \it
$^{a)}$Institut  f\"ur Kernphysik, Forschungszentrum J\"ulich,\\
D-52425 J\"ulich, Germany\medskip\\
$^{b)}$L. D. Landau Institute for Theoretical Physics,\\
GSP-1, 117940, ul. Kosygina 2, 117334 Moscow, Russia\medskip\\
$^{c)}$Department of Physics  and Mathematical Physics,\\
University of Adelaide, S.A. 5005, Australia
\vspace{1cm}\\
}
{\bf           Abstract}
\end{center}
We describe a light-cone wave function formalism for hadroproduction
of heavy-flavors via gluon-gluon fusion.
The developed approach is well suited to address
coherence effects in heavy-quark production on nuclei at small
$x_{2} \lsim 0.1\cdot A^{-1/3}$.
Nuclear attenuation in  hadroproduction of heavy-flavors is found
to be similar to shadowing effects for  heavy-flavor
structure functions in deep-inelastic scattering from  nuclei.
Nevertheless, remaining differences in the theoretical formulation of
both processes imply corrections to factorization theorems.
\pagebreak\\

\doublespace



\pagestyle{plain}

\section{Introduction}


Heavy-flavor production in hadronic interactions at high energy
is one of the  traditional applications of  perturbative Quantum
Chromodynamics (pQCD).
Although perturbative calculations are available now  in next-to-leading
order pQCD,  and seem to be in reasonable agreement with  experimental data
([1] and references therein), the production of heavy quark-antiquark ($Q\bar
 Q$)
pairs has certainly not  exceeded its
potential in revealing  precious details of strong interaction physics.
One of the missing ingredients is the quantitative understanding of nuclear
attenuation effects, which offer a possibility to learn about the propagation
of colored $Q\bar Q$ pairs through the nuclear environment.
This matter is strongly connected to effects of ``color-transparency'' and
``color-opacity'', which are  heavily discussed issues within a large variety
of
processes, ranging from quasielastic electron scattering to
heavy quarkonium production (see e.g. [2,3] and references therein).

In leading order pQCD the  cross section  for
heavy quark-antiquark  production via the collision of hadrons $h_{1}$
and $h_{2}$ at a squared center of mass energy $s$,
can be written as
\beq
\sigma_{Q\bar{Q}}(s)
=\sum_{i,j}\int dx_{1} dx_{2}\,
f_{h_{1}}^{i}(x_{1},\mu^{2})\,
f_{h_{2}}^{j}(x_{2},\mu^{2})\,
\sigma_{ij}(x_{1}x_{2}s,\mu^{2})  \, .
\label{eq:1.1}
\endeq
Here $f_{h_{1}}^{i}(x_{1},\mu^{2})$ and
$f_{h_{2}}^{j}(x_{2},\mu^{2})$ are the
densities of partons "i" and "j", carrying fractions $x_{1}$ and
$x_{2}$ of the light-cone momenta of the colliding projectile and target
 respectively.
The partonic subprocess $i+j\rightarrow Q+\bar{Q}$
is described by the cross section  $\sigma_{ij}(x_{1}x_{2}s,\mu^{2})$
and requires  a squared center of mass energy $x_1 x_2 s > 4 m_Q^2$,
where $m_Q$ is the mass of the heavy-quark.
Factorization in parton densities and hard partonic subprocesses
is carried out at a typical scale $\mu^2 \sim 4\,m_Q^2$.

In this work we will be concerned with nuclear effects in  inclusive
hadroproduction of $Q\bar{Q}$ pairs  which carry a high energy in the
laboratory
frame, where the target nucleus is at rest.
In particular  we consider processes at small target light-cone
momentum fractions  $x_2\lsim 0.1\cdot A^{-1/3}$.
Furthermore we will restrict ourselves to moderate $x_F$,
which is defined as the fraction of the projectile momentum carried by the
$Q\bar Q$ pair as seen in the center of mass frame.
Thus  we concentrate on the kinematic domain where $Q\bar{Q}$ production
is dominated by the gluon fusion subprocess $g+g\rightarrow Q+\bar{Q}$
and neither annihilation of light-quarks nor excitations of higher-twist
 intrinsic
heavy-quark components of the projectile wave function [4] are of importance.
The  pQCD cross section (\ref{eq:1.1}) is then  proportional to the density
of gluons in the beam and target.
In the case of open-charm production at  typical Tevatron energies
($s=1600\,GeV^2$)
this implies  that we focus on the region $0<x_F<0.5$.

We develop a description of heavy-flavor hadroproduction in the laboratory
frame
in terms of light-cone wave functions (LCWF) [5].
We find, that
the underlying QCD mechanism of $Q\bar{Q}$ production can  be viewed
as  diffractive  dissociation of projectile gluons into  $Q\bar{Q}$ pairs.
The resulting formalism greatly  resembles  the description of coherence
effects in diffractive scattering on nuclei.
Although there is much similarity between nuclear shadowing effects in
 hadroproduction of
heavy-flavors and shadowing of  heavy-flavor contributions
in deep-inelastic scattering (DIS), they are nevertheless different
and cannot be cast in the form of  modified nuclear gluon distributions
which are unique for  all hard scattering processes.

We apply our formalism to open-charm production and find
only small  nuclear attenuation effects in the kinematic
region considered.
For the conventional parameterization,
$\sigma_{c\bar{c}} \propto A^{\alpha}$, where $A$ is the nuclear
mass number.
Our results  correspond to an  exponent
$\alpha \approx 0.99$.
The corresponding experimental situation is not absolutely
 conclusive yet
(for a  comprehensive review on  open-charm production see
[4,6]). In early experiments [7,8] open-charm production was
estimated indirectly from the observed yield of prompt leptons,
rather than by a direct observation of produced charmed particles.
The strong nuclear effects found in these experiments,
$\alpha \sim {2\over 3}$,
led to the suggestion of large intrinsic charm components present
in the respective hadron projectile (see [4] and references therein).
More recent experiments directly  observe $D$-mesons and obtain  $\alpha
\approx 1$ , which is in good agreement with our results
($\alpha(\langle x_{F} \rangle =0.24)=0.92\pm 0.06$ in
the WA82 experiment with a 340 GeV $\pi^{-}$ beam [9];
$\alpha(0<x_{F} < 0.5)=1.00\pm 0.05$ in the E769 experiment with a
250 GeV $\pi^{\pm}$ beam [10]).

Our presentation is organized as follows: In Section 2 we analyze
the space-time picture of heavy-quark production at small $x_{2}$.
In Section 3 we develop a LCWF description for
hadroproduction of heavy-flavors on  free nucleons.
The central result in
this section is the representation of the inclusive
$Q\bar Q$ production
cross section in terms of the cross section for the scattering
of a three-particle, color singlet,
$Q$-$\bar Q$-gluon state from  the nucleon target.
We also comment on the cancellation  of comover interactions.
We extend our discussion to nuclear targets in Section 4
and apply our formalism explicitly to open-charm.
Finally we summarize and conclude in  Section 5 and indicate the
difference between nuclear attenuation in inclusive open-charm
and exclusive charmonium production.




\section{The space-time picture of $Q\bar{Q}$ production \protect\\
at small $x_{2}$}


In the laboratory frame the quantum-mechanical mechanism of nuclear
attenuation in hadroproduction of heavy $Q\bar{Q}$ pairs at small
$x_2$ and moderate $x_F$ is  similar to the mechanism of nuclear
shadowing in deep-inelastic lepton scattering at small values of the Bjorken
 variable
$x = Q^2/2M\nu$ (with $Q^2 = -\nu^2 + \vec q\,^2$ the squared momentum transfer
 and
$M$ the nucleon mass).
In DIS the absorption of the exchanged virtual photon on the  target
proceeds, at $x<0.1$, via the interaction of quark-antiquark
fluctuations present in the photon wave function.
These $q\bar q$ pairs are formed at a typical distance $\l_{q\bar q}$ from the
 target,
\beq
l_{q\bar{q}} \sim {2\nu \over Q^{2}} = {1\over M x } \,.
\label{eq:2.1}
\endeq
For $x<0.1$ the propagation length of the $q\bar q$ pairs
exceeds  the average nucleon-nucleon distance in nuclei,
$\l_{q\bar q} > R_{NN}$.
Consequently, the quark pairs will start to interact coherently with
several nucleons inside the target. This leads to shadowing (see e.g. [11,12]).
If $x$ decreases further
\beq
x \lsim {1\over R_{A}M} \sim 0.1\cdot A^{-1/3},
\label{eq:2.3}
\endeq
the propagation length of the quark pairs  becomes larger than the size of
the nuclear target itself
\beq
\lambda_{q\bar q} > R_A.
\label{eq:2.2}
\endeq
Hence, the quark-antiquark pairs will interact coherently with
the whole target nucleus while their transverse size $|\vec r|$
is frozen [11]. Shadowing will be fully developed then.

Similarly, at small values of $x_2$ and moderate $x_F$  hadroproduction
of  heavy-quarks can also be treated within the laboratory frame in terms
of $Q\bar Q$ fluctuations of gluons, interacting with the hadron projectile.
 Their typical
propagation length is
\beq
l_{Q\bar{Q}} \sim {2\nu _{G}\over 4m_{Q}^{2}} \approx
 {1\over M x_{2} } \,,
\label{eq:2.4}
\endeq
where $\nu_{G} = x_{1}E_{lab}$ denotes the laboratory energy of the
respective  parent gluon, while $E_{lab}$ stands for the beam energy.
Evidently, if the propagation length of the $Q\bar Q$ pairs
is smaller than the average  internucleon distance in the
target nucleus, $l_{Q\bar{Q}} \lsim R_{NN}$,
$Q\bar{Q}$ production on  nuclei  will be equal to
the incoherent sum over $Q\bar Q$ production on  individual
nucleons within the target.
In this regime the intranuclear
interactions of the beam gluons reduce to
rotations in color space and   the flux of gluons is preserved
\footnote{Effective nuclear attenuation is possible though
via the tension of color strings
formed by color exchanges (predominantly
on the front surface of the target nucleus) [13].}.
However, if the energy of the interacting gluons in the beam is  large
or equivalently $x_2 < 0.1\cdot A^{-1/3}$ is small,
one obtains as before
\beq
l_{Q\bar{Q}} >  R_{A}\,,
\label{eq:2.5}
\endeq
which means  that the formation of $Q\bar{Q}$ pairs takes place
coherently on the whole nucleus -- in close similarity to DIS at
small $x$.

Once produced, a  $Q\bar{Q}$ pair  evolves
into  charmed hadrons after  a typical formation (recombination)
length
\beq
l_{f} \sim {2\nu_{G} \over 4m_{D}^{2} -4m_{Q}^{2}}
\gg l_{Q\bar{Q}},
\label{eq:2.6}
\endeq
where $m_D$ is the $D$-meson mass.
The inequality between $l_{f}$ and $l_{Q\bar{Q}}$ greatly
simplifies our  further
consideration, as it implies that we may  neglect the
intrinsic evolution of a  produced
$Q\bar{Q}$ pair during its propagation through the nucleus.

At large $x_F \sim 1$ the momentum of the projectile is transferred
to a  heavy $Q\bar Q$ state collectively by several partons of the
 hadron projectile [4]. Such a  $Q\bar Q$ pair can therefore not be assigned
to a certain projectile gluon. Consequently, the excitation
of  heavy-quark components of the projectile  at $x_F \sim 1$ is quite
different
from the $Q\bar Q$ production mechanism at moderate $x_F$.
In this work we will focus on moderate $x_F$ only, where the above
mechanism is not important.

{}From a diagrammatic point of view  an accurate
analysis of coherence effects requires the calculation of
higher order diagrams, in which the incoming projectile gluon
interacts with more than one nucleon. In standard pQCD this becomes
a nearly  impossible task.
As for  DIS at $x<0.1\cdot A^{-1/3}$,
an analysis of coherence effects is greatly simplified
if one uses a LCWF  formalism.
This allows one to make explicit use of
the fact  that at high-energies, or small $x_2$, the transverse size of
partonic
fluctuations is  conserved due to Lorentz time dilatation.
A  very important difference between
hadroproduction of heavy-flavors and DIS is, that in DIS
one starts out with a color-singlet photon, whereas in the
former case the incident gluon carries  color charge. Therefore,  one
might have expected  to encounter an infrared divergent $Q\bar{Q}$
production cross section, unless the interactions with comoving spectator
 partons of
the incident color-singlet hadron were fully included.
We shall  demonstrate however, that the LCWF formalism of ref.[5] can be
generalized to hadroproduction of heavy-flavors in an infrared-stable manner.




\section{Hadroproduction of $Q\bar{Q}$ pairs on free nucleons}


In this section we develop a formalism for  hadroproduction of
heavy-flavors in terms of
LCWF valid at small $x_2<0.1$ and moderate $x_F$.
Applied to hadron-hadron collisions our results
can be shown to be equivalent to the conventional pQCD parton
model. However our new formalism
becomes truly valuable, when it is applied to
 coherence
effects in hadron-nucleus interactions.

First we will consider the production of heavy  $Q\bar Q$ pairs
through the interaction of a gluon from the projectile  with the nucleon
target. To finally obtain  the total production cross section  we have
to convolute our result with the gluon distribution of the
beam hadron.

In the LCWF technique one must carefully distinguish
between  "bare" and "dressed (physical)" partons.
In our specific process the incident
gluon will evolve after the interaction with the target
into a state which contains a $Q\bar Q$ Fock component.
If this component is projected onto a final $Q\bar Q$ state
to calculate the production amplitude of  heavy-quark pairs,
one has to bear in mind that  the physical gluon
in the LCWF formalism has a non trivial Fock decomposition which
also contains a heavy $Q\bar Q$ state.

To leading order in the QCD coupling constant, $\alpha_{S}$,
the incoming gluon $G$ has a rather simple Fock decomposition.
It  contains bare gluons, two-gluon states,
and quark-antiquark components.
As discussed in Section 2, at small $x_2<0.1$ the propagation length
of the leading order Fock states exceeds the target size.
Therefore, the $S$-matrix, describing
the interaction of the beam gluon with the target,
will not mix states containing a different number of partons.
Furthermore and most important, $S$ conserves the transverse separations  and
the longitudinal momenta of all partons present in a certain Fock state.
Consequently, it is natural to work with light-cone
wave functions in a ``mixed'' representation, given by
these conserved quantities.

In this representation  the LCWF of the physical incident gluon can be
written as (wherever it shall not lead to confusion, we suppress the
virtuality of the incident gluon $k_{1}^{2}$)
\beq
|G(k_{1}^{2})\rangle = \sqrt{1-n_{Q\bar{Q}}-n_{\xi}}|g\rangle +
\sum_{z,\vec{r}\,} \Psi_{G}(z,\vec{r}\,)|Q\bar{Q};z,\vec{r}\,\rangle
+ \sum_{\xi} \Psi(\xi)|\xi\rangle \, .
\label{eq:3.1}
\endeq
Here $\Psi_{G}(z,\vec{r}\,)$ is the projection of the physical gluon wave
 function
onto a $Q\bar Q$ state  in the  $(z,\vec{r}\,)$-representation, where  the
 light-cone
variable $z$ represents the fraction of the gluon momentum carried by the quark
and  $\vec{r}$ is the transverse separation of the $Q$--$\bar{Q}$ in the
impact parameter space.
$\sum_{\xi} \Psi(\xi)|\xi\rangle$ represents the
light quark-antiquark $q\bar{q}$ and gluon-gluon $gg$ Fock components.
The normalization of the bare gluon state in the presence of the $Q\bar Q$,
$q\bar q$ and $gg$ components is given by
\beq
\langle g|G\rangle = \sqrt{1-n_{Q\bar{Q}}-n_{\xi}}\,,
\label{eq:3.2}
\endeq
where
\beq
n_{Q\bar{Q}}= \int_{0}^{1} dz \int d^{2}\vec{r}\,
|\Psi_{G}(z,\vec{r}\,)|^{2},
\label{eq:3.3}
\endeq
determines the weight of $Q\bar{Q}$ states in the physical gluon.
Similar, $n_{\xi}$ denotes  the normalization of the light-flavor $q\bar q$ and
$gg$ components.
The $Q\bar Q$ wave function $\Psi_G$ of the gluon can be obtained
immediately from the $Q\bar Q$ wave function of a photon $\Psi_{\gamma^*}$
as derived in ref.[11].
The only difference is a color factor and the substitution
of the strong coupling constant for the electromagnetic one:
\arr
|\Psi_{G}(z,\vec{r}\,)|^{2} &=&
{\alpha_{S} \over 6\,\alpha_{em}}|\Psi_{\gamma^{*}}(z,\vec{r}\,)|^{2}
\nonumber\\
&=&{\alpha_{S}\over (2\pi)^{2}}\left\{
[z^{2}+(1-z)^{2}]\epsilon^{2}K_{1}^{2}(\epsilon r) +
m_{Q}^{2}K_{0}^{2}(\epsilon r)\right\}\,.
\label{eq:3.4}
\endarr
$K_{0,1}$ are the modified Bessel functions, $r = |\vec r\,|$,  and
\beq
\epsilon^{2}=z(1-z)k_{1}^{2} +m_{Q}^{2}.
\label{eq:3.5}
\endeq

After the interaction with the target nucleon at an  impact
parameter $\vec{b}$, the incident gluon state is transformed into
\beq
|G;\vec{b}\, \rangle \Longrightarrow  \hat{S}(\vec b\,)|G;\vec{b}\,\rangle \,.
\label{eq:3.6}
\endeq
Here $\hat{S}(\vec{b})$ is the scattering matrix in the
impact parameter representation.
The familiar eikonal form [14] of the $\hat{S}$-matrix
for the scattering of two partonic systems $a$ and $b$ is
\beq
\hat{S}(\vec b\,)=\exp\left[-i\sum_{i,j}V(\vec{b}+\vec{b}_{i}-\vec{b}_{j})
\hat{T}_{i}^{\alpha}\hat{T}_{j}^{\alpha}\right] \, ,
\label{eq:3.7}
\endeq
with the one gluon exchange potential (eikonal function)
\beq
V(\vec{b}\,)={\alpha_{S} \over \pi}\int {d^{2}\vec{k} \over
\vec{k}^{2} + \mu_{g}^{2}} \exp(i\vec{k}\cdot\vec{b}\,).
\label{eq:3.8}
\endeq
$\hat{T}_{i,j}^{\alpha}$ are color SU(3) generators
acting on the individual partons  of  $a$ and $b$ at transverse coordinates
$\vec{b}_{i}$ and $\vec{b}_{j}$ respectively.
The effective gluon mass $\mu_g$ is introduced as an infrared
regulator.
In heavy-flavor production $m_{Q} \gg \mu_{g}$,
which ensures that our  final results will
not depend  on the exact choice of $\mu_g$.
Furthermore, as we will show shortly, $\mu_g$ can be absorbed into the
definition of the target gluon density.
To the lowest non trivial order, we have to keep
terms up to  second order in $V(\vec b\,)$ only
\footnote{The fact, that the color generators in $\hat S$ do not commute
is not important to this order in $\alpha_S$,  because our final results will
contain the symmetric piece of the product of two generators only.}.

Unitarity of the $\hat{S}$-matrix at a fixed impact parameter $\vec{b}$
simplifies the calculation of the inclusive $Q\bar{Q}$ production cross
section.
For every value of $\vec{b}$ it
relates the probability $P_{Q\bar Q}$, to detect a $Q\bar Q$ pair in the final
 state
with the  probabilities $P_{G}$ and $P_{\xi}$, to find a physical  gluon
or light-flavor $q\bar q$ or $gg$ states:
\beq
P_{G}+P_{Q\bar{Q}}+P_{\xi} =1 \, .
\label{eq:3.9}
\endeq
Let us first consider $P_{G}$, which is defined through
\beq
P_{G}= {1\over 8}
\sum_{f,i,N'}
\langle G^{f}N'|\hat{S}(\vec{b})|G^{i}N\rangle
\langle G^{f}N'|\hat{S}(\vec{b})|G^{i}N\rangle^{*}\, .
\label{eq:3.10}
\endeq
$N$ and $N'$ stand for the initial target nucleon and the
final nucleonic three-quark states.
The indices "f" and "i" characterize the color
charges of the final  and incident
gluon,  where we average over the latter.
Using the explicit form of the
$\hat{S}$-matrix and applying closure for the
final nucleonic states $N'$, we
may simplify Eq.~(\ref{eq:3.10}) in order
$\alpha_S^2$ to
\beq
P_{G}= {1\over 8}  \sum_{i,f}
\langle G^{f}\bar G^{f}N|\hat{S}(\vec{b}\,)|G^{i}\bar G^{i}N\rangle \, .
\label{eq:3.11}
\endeq
We  explicitly
used the fact, that $\{-(\hat T^{\alpha})^*\}$ are the
antiparticle color generators, and
regrouped the gluon wave functions in Eq.~(\ref{eq:3.10}) accordingly.
Carrying out the color sums leaves us with the elastic scattering amplitude
for the interaction
of a color-singlet  two gluon state $(GG)_1$ with the target nucleon:
\beq
P_{G}=
\langle (GG)_{1}N|\hat{S}(\vec{b})|(GG)_{1}N\rangle \, .
\label{eq:3.12}
\endeq
Making use of the Fock state decomposition (\ref{eq:3.1}) of the
dressed  gluon,  we obtain
\arr
P_{G} &=&(1-n_{Q\bar{Q}}-n_{\xi})
\langle (gg)_{1}N|\hat{S}(\vec{b}\,)|(gg)_{1}N\rangle
\nonumber\\
 &+&2\int dz d^{2}\vec{r\,} |\Psi_{G}(z,\vec{r}\,)|^{2}
\langle (Q\bar{Q}g)_{1}N|\hat{S}(\vec{b}\,)|(Q\bar{Q}g)_{1}N\rangle
\nonumber\\
&+&2\int d\xi|\Psi(\xi)|^{2}
\langle (\xi g)_{1}N|\hat{S}(\vec{b}\,)|(\xi g)_{1}N\rangle\,.
\label{eq:3.13}
\endarr
To the considered order in pQCD  the contributions of different
flavors in (\ref{eq:3.9}) and (\ref{eq:3.13}) do not interfere.
We therefore suppress $q\bar{q}$ and $gg$ terms in the following.
In the color-singlet $(gg)_{1}$ state in Eq.~(\ref{eq:3.13}) both gluons
enter at the same impact parameter.
Consequently  this state has a vanishing color-dipole moment and
\beq
\langle (gg)_{1}N|\hat{S}(\vec{b})|(gg)_{1}N\rangle =1 \, .
\label{eq:3.14}
\endeq
\newline From Eqs.~(\ref{eq:3.13},\ref{eq:3.14}) and the
explicit form (\ref{eq:3.3}) for $n_{Q\bar{Q}}$, we obtain
\beq
P_{Q\bar{Q}} =
2\int dz d^{2}\vec{r}\,|\Psi_{G}(z,\vec{r}\,)|^{2}
\langle (Q\bar{Q}g)_{1}N|
\left[1-\hat{S}(\vec{b}\,)\right]
|(Q\bar{Q}g)_{1}N\rangle.
\label{eq:3.15}
\endeq
Note  that $P_{Q\bar Q}$ takes the form of a
profile function for the elastic scattering of a $(Q\bar Qg)_1$
state on a  nucleon. After integrating over the impact
parameter,  we finally obtain the cross section for the inclusive
production of heavy-flavor quark-antiquark pairs in gluon-nucleon
collisions:
\arr
\sigma(GN\rightarrow Q\bar{Q}\,X) &=&
2\int dz d^{2}\vec{r}\,|\Psi_{G}(z,\vec{r}\,)|^{2} \int d^{2}\vec{b}\,
\langle (Q\bar{Q}g)_{1}N|
\left[1-\hat{S}(\vec{b})\right]
|(Q\bar{Q}g)_{1}N\rangle \nonumber\\
 &=&
\int dz d^{2}\vec{r}\,|\Psi_{G}(z,\vec{r}\,)|^{2}
\sigma_{3}(\vec{r}_{1},\vec{r}_{2})
=\langle \sigma_{3N}\rangle_{Q\bar{Q}g},
\label{eq:3.16}
\endarr
where
\arr
\sigma_{3N}=
\sigma_{3}(\vec{r}_{1},\vec{r}_{2})&=&
2\int d^{2}\vec{b}\,
\langle (Q\bar{Q}g)_{1}N|
\left[1-\hat{S}(\vec{b}\,)\right]
|(Q\bar{Q}g)_{1}N\rangle \nonumber  \\
&=&
{9\over8}\left[\sigma(r_{1})+\sigma(r_{2})-
{1\over 9}\sigma(r)\right]\,.
\label{eq:3.17}
\endarr
As indicated  $\sigma_{3N}$ is the
total interaction cross section for the scattering of a
color-singlet three-parton state $(Q\bar{Q}g)_1$ from a nucleon [5].
The  $g$-$Q$ and $g$-$\bar{Q}$ separation  in the impact parameter
space are  labeled $\vec{r}_{1}$ and $\vec{r}_{2}$  respectively.
Their difference $\vec{r}=\vec{r}_{1}-\vec{r}_{2}$
is identical to the  $Q$-$\bar{Q}$ separation.
The distances $\vec{r}_{1,2}$ are related  to the light-cone
momentum fractions carried by the $Q$ and $\bar{Q}$  respectively:
\beq
\vec{r}_{2}=- z\vec{r},~~\vec{r}_{1}=(1-z)\vec{r}\, .
\label{eq:3.18}
\endeq
The cross section $\sigma (r)$ for the interaction of a $Q\bar Q$
color-dipole of
size $r=|\vec r\,|$ with a nucleon reads in Born approximation [11]
\beq
\sigma_B(r) =
{16 \over 3}\int {d^{2}\vec{k} \over (\vec{k}^{2}+\mu_{g}^{2})^{2}}
\alpha_{S}^{2}\, \left(1-{\cal F}_{2}(\vec{k})\right)
\,\left(1-\exp(i\vec{k}\cdot\vec{r}\,)\right)\, ,
\label{eq:3.19}
\endeq
where ${\cal F}_{2}(\vec{k})=\langle N|
\exp(i\vec{k}\cdot (\vec{\rho}_{1}-\vec{\rho}_{2}))|N\rangle$
is the two-quark formfactor of the nucleon.
In general, the dipole cross section is related in the
Leading-Log($Q^2$) approximation (LLQA) to the gluon distribution
of the target [5,16,17]:
\beq
\sigma(r) \rightarrow \sigma(x_{2},r)={\pi^{2} \over 3}\,r^{2}\,
\alpha_{S}(r)
\left[
x_{2}\,g(x_{2},k_{2}^{2} \approx {{\cal A}\over r^{2}})\right],
\label{eq:3.20}
\endeq
where ${\cal A}\approx 10$ [18].
The explicit $x_2$ dependence in (\ref{eq:3.20}) results from
higher order Fock components of the incident physical gluon,
e.g. $Q\bar Q g$ states, and is determined  through  the BFKL evolution
equation (for a  detailed BFKL phenomenology of the dipole cross section
see [15]).
The gluon distribution in (\ref{eq:3.20}) absorbs the  dependence
on the infrared regulatization $\mu_g$ in the same manner as in
the conventional QCD analysis of hard scattering processes.
The most important point for our further discussion
is the color transparency property, i.e. the $\propto r^{2}$
dependence of the dipole cross section in (\ref{eq:3.20}).

When cast in the form (\ref{eq:3.16}), the
heavy-flavor production cross section
resembles heavy-flavour contributions to  real and virtual
photoproduction.
While the three particle cross section (\ref{eq:3.17})
enters in the former, the dipole cross section $\sigma(r)$ is present in the
 latter.
(See also Eq.~(\ref{eq:3.4}) for the relationship between the corresponding
wave functions.)
Both processes have a similar
dipole-size structure, and we briefly recapitulate the
results of refs.[5,11]:
The $Q\bar Q$ wave function (\ref{eq:3.4}) decreases exponentially
for $r \gsim 1/\epsilon$. We may therefore conclude  that for
$k_{1}^{2} \lsim 4m_{Q}^{2}$
small transverse sizes
are relevant for the  $Q\bar Q$ production process:
\beq
r^{2}, r_{1}^{2},r_{2}^{2} \lsim {1\over m_{Q}^{2}} \, .
\label{eq:3.21}
\endeq
For highly virtual gluons, $k_{1}^{2} \gg 4m_{Q}^{2}$, the
$Q\bar{Q}$ production cross section (\ref{eq:3.16})
vanishes $\sim 1/k_1^2$ due to the color transparency
property of the dipole cross section. (In view of the analogy
to high energy photoproduction, this behavior can be
understood as the counterpart of Bjorken scaling.)
Combining equations (\ref{eq:3.16},\ref{eq:3.20}, \ref{eq:3.21}), we therefore
 find
\beq
\sigma(GN\rightarrow Q\bar{Q}X) \propto x_{2}g(x_{2},\mu^{2}
\sim 4m_{Q}^{2})\, .
\label{eq:3.22}
\endeq

The color transparency property (\ref{eq:3.20}) of the dipole
cross section has important implications.
First note,  that $n_{Q\bar{Q}}$, the probability
that a physical gluon is realized in a $Q\bar{Q}$ Fock component, has
a familiar logarithmic ultraviolet-divergence at
$r\rightarrow 0$.
Nevertheless, in the LCWF formalism the heavy-quark
production cross section
$\sigma(GN \rightarrow Q\bar Q\,X)$ is ultraviolet finite, since
the color transparency property of the dipole
cross section (\ref{eq:3.20}) leads to a rapid
convergence of the $r$-integration in the cross section
(\ref{eq:3.16}).
The physical reason  for color transparency to play a role
in this process is, that the small size $Q\bar Q$ pairs
cannot be resolved by  interacting t-channel gluons with
wavelengths $\lambda \gsim r$. They are therefore indistinguishable
from bare gluons and cannot be excited into $Q\bar Q$ states.

The second remarkable feature  of the LCWF formalism  is  that
$\sigma(GN \rightarrow Q\bar Q\,X)$ is also  infrared-stable.
This is not the case for the total gluon-nucleon cross section,
being  (logarithmically) infrared-divergent for  $\mu_{g}\rightarrow 0$.
Only the interaction of two color-singlet hadrons
yields an  infrared-finite total cross section.
This fact is due to the cancellation of the incident gluon color charge
through  the color charge of comoving spectator partons belonging to the
 color-singlet
hadron projectile.
However, soft t-channel gluons with $\vec k \rightarrow 0$ cannot
resolve the $Q\bar Q$ component of the physical beam gluon and
therefore cannot contribute to heavy-flavor production.
We consequently obtain an infrared-stable
$Q\bar Q$ production cross section (\ref{eq:3.16})
already for an isolated, colored, gluon projectile.

A simple consideration shows, that the $Q\bar Q$
production cross section
(\ref{eq:3.16}) is not affected by the presence
of spectator partons in the
 beam.
Consider for example the case of one spectator "parton" $s$,
which complements the incident gluon to a
color-singlet state $(Gs)_1$.
The spectator parton is assumed to scatter elastically only, i.e.
without dissociating  into a $Q\bar{Q}$ pair.
Generalizing the above analysis, we finally end up with
\beq
P_{G} \propto \sum_{f,d}
\langle (Gs)_1 (\bar G\bar s)_1 \, N |\hat{S}(\vec{b}\,)|
                (G^{f}s^{d}) (\bar G^f \bar s^d) \,N \rangle \,.
\label{eq:3.23}
\endeq
To obtain the inclusive production cross section  we have to  sum over
the colors ``$f$'' and ``$d$'' of the final gluon and spectator states
respectively. This yields a color-singlet two gluon state, as well as
a color-singlet spectator $s\bar s$ state.
Since both partons $s$  and $\bar s$ of this singlet state enter at the
same impact parameter, their color-dipole moment vanishes and they
decouple  from the interacting gluons.
This leaves us  with our previous result given in Eq.~(\ref{eq:3.16}).
A similar cancellation
takes place for an  arbitrary number of spectators,
as long as we neglect
interference effects between $Q\bar{Q}$ pairs belonging
to different gluons of the beam hadron.
However such a restriction to elastic rescatterings of  spectators
is a standard ingredient of  pQCD calculations in the
framework of the conventional LLQA.
In this respect the cancellation  of interactions
with spectator partons
can be traced back to the LLQA factorization of
the LCWF, containing
the heavy-quark component.
Consequently in LLQA, the production cross section of $Q\bar{Q}$
pairs in the kinematic region under  consideration is determined by
the gluonic subprocess
$G+N\rightarrow Q\bar{Q}+X$, i.e. the underlying process
is  diffractive dissociation of  gluons into $Q\bar Q$ pairs.

We are now in the position to write down the heavy-quark production
cross section for hadron-nucleon collisions $\sigma_{Q\bar Q}(h,N)$.
For this purpose we have to
multiply $\sigma(GN\rightarrow Q\bar Q\,X)$ from
Eq.~(\ref{eq:3.16}) with the gluon density of the incoming hadron
projectile
and integrate over the virtuality $k_1^2$ of the projectile gluon:
\beq
{d\sigma_{Q\bar{Q}}(h,N) \over dx_1}=
\int {dk_{1}^{2} \over k_{1}^{2}}
{\partial[g_{1}(x_{1},k_{1}^{2})]
 \over \partial \log k_{1}^{2}}\,
 \sigma(GN \rightarrow Q\bar{Q}\,X;x_2,k_{1}^{2}).
\label{eq:3.24}
\endeq
As mentioned above, the leading contributions to
$\sigma(GN\rightarrow Q\bar Q\,X)$ result from the region
$k_1^2 \lsim 4 m_Q^2$, where we  approximate
$\sigma(GN \rightarrow Q\bar Q\,X;x_2,k_{1}^{2}) \approx
\sigma(GN\rightarrow Q\bar Q\,X;x_2,k_{1}^{2}=0)$.
We therefore obtain
\beq
{d\sigma_{Q\bar{Q}}(h,N) \over dx_{1}}\approx
g_{1}(x_{1},\mu^{2} = 4m_{Q}^{2})\,
\sigma(GN\rightarrow Q\bar{Q}\,X;x_{2},k_{1}^{2} = 0) \,.
\label{eq:3.25}
\endeq
Notice, that we have recovered the dependence on the gluon density of
the beam as in the parton model ansatz (\ref{eq:1.1}), whereas
(\ref{eq:3.22}) yields  the gluon density of the target.
This shows the equivalence of the LCWF formulation of
heavy-flavour production and the conventional parton model approach.

Let us explore the above result for charm production in nucleon-nucleon
collisions. To be specific we choose a beam energy
$E_{lab} = 800$\,GeV,
as used in the Fermilab E743 experiment [19].
We employ the  dipole cross section of ref.~[15], which has been
shown to yield a good quantitative description of the closely
related real photoproduction of open-charm, and
the small $x$ proton structure function, as measured at  HERA.
According to the analyses in refs.~[18,20],
one obtains for transverse sizes $r^2 \sim 1/m_c^2$ a
precocious asymtotic BFKL behavior
\beq
\sigma(x_{2},r) \approx \sigma_{B}(r)
\left({x_{0}\over x_{2}}\right)^{\Delta}\, ,
\label{eq:3.26}
\endeq
valid at $x_{2} \leq x_{0}$.
The Born cross section $\sigma_{B}(r)$ from Eq.~(\ref{eq:3.19})
is evaluated for $\mu_{g}=0.75$\,GeV, $\Delta = 0.4$ is the
intercept of the BFKL pomeron and $x_{0}=0.03$.
The charm quark mass is fixed at $m_c = 1.5$\,GeV.
The gluon distribution of the nucleon projectile enters at
moderate to large values of $x_{1}$. We may therefore use the
parameterization of  ref.[21].
Because the $g+g\rightarrow Q+\bar{Q}$ cross section
decreases rapidly with the invariant mass of the heavy-quark pair, one finds
$x_1 x_2 s \sim 4 m_Q^2$.
We may now calculate
$\frac{d\sigma_{c\bar c}(N,N)}{dx_F}$ from Eq.~(\ref{eq:3.25}).
In Fig.1 we compare our results with the
data from the  E743 proton-proton
experiment [19].
In the kinematic region, where our approach is well founded
$0<x_F<0.5$,  we obtain resonable
agreement with the experimental data.




\section{Hadroproduction of $Q\bar{Q}$ pairs on nuclear targets}


Although the conventional pQCD result for heavy-flavor production
and the LCWF formulation are equivalent, there is an important
advantage in the latter. Namely in equation (\ref{eq:3.16})
for $\sigma (GN\rightarrow Q\bar Q\,X)$
the LCWF  of the $Q\bar Q$ state and its interaction
cross section factorize.
This feature is crucial for the now following generalization
to  nuclear targets.

The derivation of Eq.~(\ref{eq:3.16})
was based upon the observation that
at high energies, where the coherence length of the  $Q\bar Q$
fluctuation of the  incident
gluon is larger than the target size
$l_{Q\bar{Q}} > R_{target}$,  the
transverse separations of the beam partons
are  frozen. This led to a diagonalization
of the scattering $\hat{S}$-matrix
in a mixed $(z,\vec{r}\,)$-representation.
In high energy hadron-nucleus collisions
with $l_{Q\bar{Q}} \gsim R_{A}$,
this diagonalization of the $S$-matrix
can  evidently be achieved too.
Consequently we obtain the $Q\bar Q$
production cross section for nuclear
targets by substituting  $\sigma_{3A}$ for $\sigma_{3N}$
in  Eq.~(\ref{eq:3.16}).
The cross section $\sigma_{3A}$  for the scattering of a
three-parton, color singlet, $(Q\bar Q g)_1$ state from
a nucleus with mass number $A$  is given in
the frozen size approximation  through the
conventional Glauber formalism (we suppress the parton separations
$\vec{r},\vec{r}_{1,2}$):
\beq
\sigma_{3A}=
2\int d^{2}\vec{b} \left\{
1-\left[1-{1\over 2A}\
\sigma_{3N}T(\vec b\,)\right]^{A}\right\} \nonumber
\approx
2\int d^{2}\vec{b} \left\{
1-\exp\left[-{1\over 2}\
\sigma_{3N}T(\vec b\,)\right]\right\}  \, .
\label{eq:4.1}
\endeq
Here $\vec{b}$ is the impact parameter of  the $(Q\bar{Q}g)$-nucleus scattering
 process,
which must not be confused with the impact parameter of the
 $(Q\bar{Q}g)$-nucleon
interaction in Section 3. $T(\vec b\,)$ stands for the optical thickness of the
 nucleus
\beq
T(\vec b\,)=\int_{-\infty}^{+\infty} dz \,n_{A}(\vec b,z) \,,
\label{eq:4.2}
\endeq
with the nuclear density ${n_A}(\vec b,z)$ normalized to
$\int d^{3}\vec{r} \,n_{A}(\vec r\,)=A$.
In LLQA the  inclusive cross section for $Q\bar{Q}$
hadroproduction on nuclei  can  then be written as
\beq
{d\sigma_{Q\bar{Q}}(h,A) \over dx_{1}} \approx
g_{1}(x_{1},\mu^{2} = 4m_{Q}^{2})\,
\sigma(GA\rightarrow Q\bar{Q}\,X;x_2,k_{1}^{2}=0) \, ,
\label{eq:4.3}
\endeq
where
\begin{eqnarray}
&&\sigma(GA\rightarrow Q\bar{Q}\,X;x_2,k_{1}^{2}) =
\nonumber\\
&&\qquad2\,\int dz d^{2}\vec{r}\,|\Psi_{G}(z,r)|^{2}
\int d^{2}\vec{b} \left\{
1-\exp\left[-{1\over 2}\
\sigma_{3N}T(\vec b\,)\right]\right\} =
\langle \sigma_{3A}\rangle_{Q\bar{Q}g}
\, .
\label{eq:4.4}
\end{eqnarray}
In the multiple scattering series (\ref{eq:4.1},\ref{eq:4.4})
nuclear coherence effects are controlled by the expansion parameter
\beq
\tau_{A}=\sigma_{3N}\,T(\vec b\,)
\propto \sigma_{3N}\,A^{1/3}.
\label{eq:4.5}
\endeq
Expanding the exponential in (\ref{eq:4.1},\ref{eq:4.4})
in powers of $\tau_{A}$ one can identify  terms
$\propto \tau_{A}^{\nu}$. They describe contributions to the
total production cross section,  resulting from  the coherent
interaction of the $(Q\bar Qg)$ state with $\nu$ nucleons inside the
target nucleus.
In leading order, $\nu = 1$,
 we remain with the incoherent sum over
the nucleon production cross sections:
\arr
\sigma(GA\rightarrow Q\bar{Q}\,X) &=&
\int d^{2}\vec{b}\,
T(\vec b\,)\,
\sigma(GN\rightarrow Q\bar{Q}\,X) \nonumber \\
&=&A\,\sigma(GN\rightarrow Q\bar{Q}\,X).
\label{eq:4.6}
\endarr
This yields the conventional impulse approximation  component
of the nuclear cross section, which is proportional to the
nuclear mass number.
Effects of coherent higher  order interactions are  usually discussed
in terms of the nuclear transparency
\beq
T_{A}= {\sigma_{Q\bar Q}(h,A) \over {A\, \sigma_{Q\bar Q}(h,N)} } \, .
\label{eq:4.7}
\endeq
In the impulse approximation  $T_{A} = 1$. The driving contribution to
nuclear attenuation results  from the coherent interaction of the
three-parton $(Q\bar{Q}g)$ state with two nucleons inside the target:
\beq
T_{A}=1-{1\over 4} \,
{\langle \sigma_{3N}^{2} \rangle_{Q\bar{Q}g}
\over
\langle \sigma_{3N}\rangle_{Q\bar{Q}g} }\int d^{2}\vec{b}\,T^{2}(\vec b\,)
\, .
\label{eq:4.8}
\endeq
At $k_{1} ^{2} \ll 4m_{Q}^{2}$   shadowing effects can easily
be estimated, using the fact  that
$|\Psi_{G}(z,r)|^{2} \propto \exp(-2m_{Q}r)$ and $\sigma(r) \approx
Cr^{2}$.
We further  approximate
$\sigma_{3}(\vec{r}_{1},\vec{r}_{2})\approx {1\over 2}\sigma(r)$,
and obtain
\beq
{\langle \sigma_{3N}^{2} \rangle_{Q\bar{Q}g}
\over
\langle \sigma_{3N}\rangle_{Q\bar{Q}g} }
\approx {5C \over 2m_{Q}^{2}} \, .
\label{eq:4.9}
\endeq
For Gaussian nuclear densities with mean square charge radii $R_{ch} = 1.1
 A^{1/3}\,fm$
we find
\beq
\int d^{2}\vec{b}\,T^{2}(\vec b\,) =  {3A \over 4\pi R_{ch}^{2}}\,.
\label{eq:4.10}
\endeq
Altogether, we finally obtain the following estimate for nuclear attenuation
effects:
\beq
1-T_{A}\approx
{15CA \over 32\pi R_{ch}^{2} m_{Q}^{2}}\,.
\label{eq:4.11}
\endeq
At pre-asymptotic energies, when the coherence length $l_{Q\bar{Q}}
\lsim R_{A}$, the shadowing term in (\ref{eq:4.8})
will be proportional to the squared
nuclear charge form factor  $F_{A}^{2}(\kappa)$, where
$\kappa = 1/l_{Q\bar{Q}}=Mx_{2}$ is the longitudinal momentum
transfer in the $G+N\rightarrow Q\bar{Q}+N$ transition [22]. This
form factor quantifies the onset of nuclear coherence  effects.

Applied to open-charm production at current Tevatron energies
($s=1600\,GeV^2$) we have to remember that $x_2$ is bound from
below through the kinematic  constraint
\beq
x_{1}x_{2} \gsim {4m_{c}^{2} \over s} \approx (0.5-1)\cdot 10^{-2}.
\label{beq:4.10}
\endeq
In the accessible region of small $x_2 \sim 0.01$ we find at
$r\sim 1/m_c$ typically $C \approx  2.5$ [15], which leads to
\beq
1-T_{A} \sim 6\cdot 10^{-3}\cdot A^{1/3}\, .
\label{eq:4.12}
\endeq
Approximated by the ansatz $T_{A} \approx A^{\alpha-1}$, we get
$1-\alpha \sim 7\cdot 10^{-3}$, in agreement with more recent experimental
results [9,10].
However,  this also shows that data with  high accuracy
are necessary to observe nuclear attenuation of open-charm production.

Although nuclear shadowing effects turn out to be
small, it is interesting to consider  their consequences for
factorization theorems.
Remember the close similarity between  nuclear shadowing
in heavy-flavor hadroproduction and heavy-flavor
contributions to nuclear structure functions.
It mainly involves an exchange of the three-parton  cross section $\sigma_{3N}$
in (\ref{eq:4.8}) with the dipole cross section
$\sigma(r)$, which differs  in the considered process  by
a  factor $\sim 2$ (see the discussion following
Eq.~(\ref{eq:4.8})).
Indeed,  comparable $\sim 1\%$ nuclear shadowing
for the charm structure function of lead was found in ref.[16].
Following the analysis of ref.[11] we note, that the
shadowing effect in  Eq.(\ref{eq:4.8}) does not vanish at high
virtualities  of the projectile gluon $k_{1}^{2} \gg 4m_{Q}^{2}$,  and
consequently  is a  leading twist effect. (The same is true for
nuclear shadowing in DIS.)
Our formalism correctly reproduces the proportionality
of the  $Q\bar{Q}$ hadroproduction
cross section to the gluon structure function of
the target nucleon, in agreement with the conventional parton
model formulation as shown in Section 3.
A similar proportionality to the gluon
structure function of the target nucleon holds also for
real (and weakly virtual $Q^{2} \lsim 4m_{c}^{2}$) photoproduction
of  open-charm on nucleons. It is therefore tempting to assume such a
factorization also in nuclear production processes.
However the three-parton cross section and the dipole cross section,
entering hadro- and photoproduction respectively,  are not equal.
Hence  the $A$-dependence of both processes cannot be
described in terms  of  modified nuclear  gluon distributions,
which would be unique for all hard processes.
The fact,  that nuclear shadowing defies  factorization,
despite being a leading twist effect,
was already encountered earlier in connection with
DIS and Drell-Yan production on nuclei [16],  and
also in  nuclear production of vector mesons [23] and jets [24].

A detailed discussion of the subtleties of  factorization
breaking goes beyond the scope of the present paper. Here we only
wish to note, that in DIS  and  heavy-flavor
hadroproduction  non-factorizable effects result from the contributions of
$Q\bar{Q}$ Fock states  present in the projectile photon or gluon respectively.
Furthermore, shadowing of $Q\bar Q$ contributions in
DIS must be reinterpreted
as a modification of the nuclear sea quark distributions  [5]. This, however,
is not possible for heavy-flavor hadroproduction.
Shadowing effects, which can be ascribed
to nuclear modifications of gluon structure
functions,  emerge in DIS first through  contributions from higher order
$Q\bar{Q}g$ Fock states of the photon.
For the scattering from free nucleons these lead to the conventional
QCD evolution of the target gluon distribution function [5] included in
Eq.~(\ref{eq:3.20}).
However in the case of nuclear targets the situation becomes different as we
will  outline briefly.
In DIS at large momentum transfers  a $Q\bar Qg$ Fock state, which contains a
 soft
gluon, is characterized through a small transverse separation of the
 quark-antiquark pair
$r_{Q\bar{Q}} \sim 2/\sqrt{Q^{2}+4m_{Q}^{2}}$, but a large separation
$r \sim R_{g} ={1/\mu_{g}}$ of the gluon from both quarks (see refs.[5,25]).
Note that the latter is independent of the photon  virtuality
$Q^2$ and  $m_Q$.
A $Q\bar Qg$ Fock state of the virtual photon can therefore be treated
as an octet-octet color dipole of size $r$, because the $Q\bar Q$ pair
acts as a pointlike color charge by virtue of $r \gg r_{Q\bar Q}$.
Furthermore the squared
wave function of the $Q\bar{Q}g$ Fock state factorizes into
the square of the $Q\bar{Q}$ wave function
$|\Psi_{G}(z,r_{Q\bar{Q}})|^{2}$, and the distribution of soft
gluons around the nearly pointlike $Q\bar{Q}$ pair:
\beq
|\Psi_{3}(z,r_{Q\bar{Q}},z_{g},r)|^{2}=
{1\over z_{g} }\cdot {4\over 3\pi}\alpha_{S}(r_{Q\bar{Q}})
|\Psi_{\gamma^{*}}(z,r_{Q\bar{Q}})|^{2}
{r_{Q\bar{Q}}^{2}\over r^{4}}\Phi(\mu_{g}r)\,.
\label{eq:4.13}
\endeq
Here $z_{g}$ is the momentum fraction  carried by the soft
gluon and $\Phi(x)=x^{2}[K_{1}^{2}(x)+xK_{1}(x)K_{0}(x)+
{1\over 2}x^{2}K_{0}^{2}(x)]$.
Nuclear attenuation of  $Q\bar{Q}g$ states will be controlled
by the color octet dipole cross section $\sigma_{8}(r)={9\over 4}\sigma(r)$
[5].
On the other hand, the contributions of  $Q\bar{Q}g$ components to
shadowing can be related to diffractive dissociation of virtual photons
into large mass states $M^2 \gg Q^2$, which is successfully
described in terms of the so-called triple-pomeron coupling $A_{3\Pom}$.
At $x_{2} \lsim x_A = 0.1 \,A^{1/3}$ one  finds  [5,22,26]
\beq
1-T_{A} \approx
{3A\over R_{ch}^{2}} A_{3\Pom}\log\left({x_{A}\over x}\right)
\sim 1.5\cdot 10^{-2}A^{1/3}\log\left({x_{A}\over x}\right)\,.
\label{eq:4.14}
\endeq
We used $A_{3\Pom}\approx 0.16$\,GeV$^{-2}$ as determined
in the photoproduction experiment [27].

Similar considerations are applicable to the excitation of
$Q\bar{Q}g$ states in  hadroproduction of heavy-flavors.
Postponing   detailed numerical calculations,
we may only cite the main ideas:
The underlying process will be  diffractive excitation of  incident
projectile gluons $G_{1}$ into quark-antiquark-gluon states
$G_{1}\rightarrow Q\bar{Q}g_{2}$.
Consequently the generalization of the nucleon (\ref{eq:3.16})
and nuclear (\ref{eq:4.4}) production cross section will contain the
interaction cross section of a  four-parton color singlet state
$Q\bar{Q}g_{1}g_{2}$, where  the transverse size $r_{Q\bar{Q}} \sim 1/m_Q$
of the $Q\bar{Q}g_{1}$ component is much smaller than the separation
$r\sim R_{g}$ of the soft gluon $g_{2}$.
Furthermore they will involve
the  factorized three-parton wave function from  Eq.~(\ref{eq:4.13}).
Since the size of the three-parton $Q\bar Q g$ state in DIS,
as well as the size of the four-parton $Q\bar Q g_1 g_2$ state in
hadroproduction is  determined by the soft gluon separation $r\sim 1/R_g$,
nuclear attenuation of $Q\bar Q g$ states will be similar in both processes.
Only $x_2$ has to be substituted for  $x$ in Eq.~(\ref{eq:4.14}).

At $R_{g}\gsim r_{Q\bar{Q}}$ the triple pomeron coupling becomes
approximately independent of $Q^2$ and quark masses [5].
Therefore  we may
reinterpret the nuclear attenuation of $Q\bar Qg$ Fock states as
nuclear modifications of target gluon distributions. These are then
unique for  all hard processes.
However in DIS the non-factorizable contributions of light flavor
$q\bar q$ pairs dominate nuclear shadowing down to
$x \lsim 10^{-3}$. Higher order Fock states are important at
much smaller $x$ only [5].
Hence the observed nuclear shadowing in DIS is not an appropriate
place to determine nuclear modifications of gluon distributions.
In open-charm hadroproduction the triple-pomeron contribution
(\ref{eq:4.14}) will be relevant already at $x_{2}\sim 10^{-2}$.
Nevertheless possible attenuation effects will
not exceed $\sim$ (5-10)\%  even for heavy nuclei.
This corresponds to an exponent
$0.98 \lsim \alpha \lsim 1$ -- a deviation from unity
which is still below the current experimental accuracy.
Of course Eq.~(\ref{eq:4.14}) gives a crude estimate
of the $x_{2}$ dependence of nuclear attenuation only. A
more detailed calculation using the pomeron structure function of ref.~[25]
will be presented  elsewhere.

Finally, the estimate (\ref{eq:4.14}) shall be applicable to nuclear shadowing
in Drell-Yan and exclusive vector meson production on nuclei.
In the latter case  $q\bar{q}$ contributions
to nuclear shadowing decrease
 asymptotically
$\propto 1/(Q^{2}+m_{V}^{2})$, but are  numerically very large.
The influence of  color transparency [28] demands
an extremely large momentum transfer $Q^{2}$ for a substantial
decrease of quark-antiquark $q\bar{q}$ contributions and the onset of the
universal attenuation of Eq.~(\ref{eq:4.14}).


\section{Conclusions}


The purpose of this paper was  the  presentation  of
a light-cone wave function approach to hadroproduction of heavy-quarks
on nucleons and nuclei  valid at small $x_2 \lsim 0.1\,A^{1/3}$
and moderate $x_F\not\sim 1$.
We found that in the laboratory frame  the mechanism of heavy-flavor production
 in
LLQA can be described as a diffractive dissociation of projectile gluons into
$Q\bar{Q}$ pairs.
The central result is the representation of the inclusive
$Q\bar Q$ production
cross section in terms of the cross section for the scattering
of a three-parton, color singlet,
$Q$-$\bar Q$-gluon state from the target.
Although the incident gluon carries color charge,
the light-cone formalism leads to an infrared-stable
heavy-flavor production cross section, which
is not affected by interactions with
comoving spectators.
Nuclear attenuation effects were found  to be small
$\sim A/(m_Q R_A)^2$.
We applied our formalism to open-charm production and obtained
good agreement with more recent experimental data.
Comparing nuclear effects in photoproduction and hadroproduction
of open-charm  we observed,  that despite beeing a leading twist effect,
leading shadowing  effects defy factorization.
However nuclear attenuation of higher order production processes can
be cast in the form of nuclear modifications of gluon distributions.

Weak nuclear attenuation in inclusive hadroproduction of open-charm
does not imply weak nuclear effects for  exclusive charmonium
channels.
It should however be possible to extend the presented technique to
the latter.
As far as nuclear effects are concerned, the difference between both
should  be quite similar  to the well established
difference between inclusive photoproduction of open-charm and exclusive
photoproduction of charmonium bound states.
In the first case nuclear attenuation is controlled by the
$c\bar c$ dipole cross section at small distances $r\sim {1/ m_{c}}$.
This leads to a weak nuclear attenuation [5,16].
On the other hand in  exclusive production processes the typically transverse
scale is given by the size of the
charmonium bound states $R_{J/\Psi},R_{\chi}
\gg {1/ m_{c}}$, with the consequence
of much stronger  nuclear effects [28].
A detailed treatment of exclusive
charmonium production will however be
presented elsewhere.

Finalizing this paper we learned  that
B. Z. Kopeliovich is working on a similar analysis of nuclear shadowing
effects in  Drell-Yan processes [29].

\bigskip
\bigskip

{\bf Acknowledgments:} G. P. and B. G. Z.
would like to thank J. Speth and the theoretical physics group at
the Institut f\"ur Kernphysik, KFA J\"ulich, for their hospitality
during several visits. B. G. Z. acknowledges discussions with
B. Z. Kopeliovich.
We also thank A. W. Thomas for a careful reading of the
manuscript.
This work was partially supported by the INTAS grant 93-239 and
grant N9S000 from the International Science Foundation.
\pagebreak\\

{\bf \Large Figure caption:}
\begin{itemize}

\item[Fig.1]
     Differential cross section of
     open-charm production in nucleon-nucleon
     collisions at $E_{lab}=800$\,GeV. The experimental
     data are from ref.[14].

\end{itemize}
\end{document}